\documentclass[twocolumn]{aastex631}
\begin{document}

\title{Blazar PKS 0446+11 - Neutrino connection study using a lepto-hadronic model}

\author[0000-0002-6941-7002]{Rukaiya Khatoon}
\affiliation{Centre for Space Research,
North-West University, 
Potchefstroom, 2520, South Africa}

\author[0000-0002-8434-5692]{Markus B\"ottcher}
\affiliation{Centre for Space Research,
North-West University,
Potchefstroom, 2520, South Africa}

\author[0000-0001-6399-3001]{Joshua Robinson}
\affiliation{Centre for Space Research,
North-West University,
Potchefstroom, 2520, South Africa}


\begin{abstract}
We present a multi-wavelength study of a blazar PKS 0446+11, motivated by its spatial association with the neutrino event IC240105A detected by the IceCube Neutrino Observatory on 2024 January 5. The source is located 0.4$^{\circ}$ from the best-fit neutrino direction and satisfies selection criteria for VLBI-selected, radio-bright AGN that have been identified as highly probable neutrino associations. PKS 0446+11 exhibited a major $\gamma$-ray flare in November 2023, reaching $\sim 18 \times$ its 4FGL-DR4 catalog average. Around the neutrino epoch, PKS 0446+11 remained in an elevated state, with the $\gamma$-ray flux more than six times above its catalog level, the X-ray flux an order of magnitude above the archival measurements, and the optical-UV emission also enhanced. We used Fermi-LAT, Swift-XRT/UVOT, and archival multi-wavelength data to construct multi-wavelength light curves and spectral energy distributions (SEDs). SED modeling shows that the emission is best described by a leptonic scenario, with synchrotron emission at low energies and external Compton scattering of broad-line region and dusty torus photons dominating the X-ray–$\gamma$-ray output.
 A lepto-hadronic model fails to adequately reproduce the observed SED, although hadronic cascades can broadly account for the X-ray and $\gamma$-ray spectral coverage at lower flux levels. We compute the expected neutrino flux for the hadronic scenario and compare it to the IceCube 90\% upper limit. Our results highlight the importance of continued multi-wavelength and neutrino monitoring to better understand the physical conditions under which this blazar may serve as neutrino source.
\end{abstract}


\section{Introduction} \label{sec:intro}

The discovery of a diffuse flux of high-energy astrophysical neutrinos by the IceCube Neutrino Observatory 
(\citealt{Aartsen_2013,Aartsen_2014}) has opened a new era in multi-messenger astronomy, providing unique insights into the most powerful particle accelerators in the Universe. Neutrinos, unlike photons, travel essentially unimpeded through cosmic distances, unaffected by intervening matter, radiation fields, or magnetic deflections. Their detection therefore provides a direct probe of hadronic processes in astrophysical sources (\citealt{Ahlers2015,Murase2016}), offering unambiguous evidence of the acceleration of protons or heavier nuclei to ultra-relativistic energies. In high-energy environments, such as those found in relativistic jets or compact accretion flows, cosmic rays can interact with ambient matter ($pp$ interactions) or with photon fields ($p\gamma$ interactions), producing charged and neutral pions. The decay of charged pions leads to the production of muons and neutrinos, while neutral pions decay into high-energy gamma rays. This naturally links neutrino production to gamma-ray emission, although the relationship can be complicated by the different interaction cross-sections, energy thresholds, and attenuation processes affecting photons, but not neutrinos.

Among the known astrophysical classes, active galactic nuclei (AGN) with relativistic jets, in particular blazars, are among the most powerful and efficient persistent particle accelerators. In blazars, the jet is closely aligned with our line of sight, causing the emission to be strongly Doppler-boosted and highly variable across the electromagnetic spectrum. Blazars are conventionally divided into flat-spectrum radio quasars (FSRQs), which exhibit strong, broad emission lines in their optical spectra, and BL Lacertae objects (BL Lacs), whose optical spectra are dominated by a featureless, nonthermal continuum with weak or absent emission lines \citep{Stickel1991}. Both subclasses display broadband spectral energy distributions (SEDs) characterized by two broad components: a low-energy hump extending from radio through optical/UV and in some cases into the X-ray band,  and a high-energy hump spanning X-rays to GeV–TeV $\gamma$-rays \citep{Abdo2010,Bottcher2013}. 
The low-energy hump is generally attributed to synchrotron emission from relativistic electrons gyrating in the jet’s magnetic field, while the origin of the high-energy hump remains debated.

In purely leptonic scenarios, 
the high-energy hump arises from inverse-Compton (IC) scattering. The seed photons can be the synchrotron photons themselves, giving rise to the synchrotron self-Compton (SSC) process, or can be external to the jet, such as thermal emission from the accretion disk, reprocessed photons from the broad-line region (BLR), or infrared emission from a dusty molecular torus (external Compton; EC) \citep{Dermer1992,Sikora1994,Ghisellini1998} — collectively referred to as external Compton (EC) processes. These models have been successful in reproducing the broadband emission of many blazars, especially during quiescent or moderately active states.

However, in lepto-hadronic or purely hadronic scenarios, high-energy emission can arise from ultrarelativistic protons, which may radiate through proton synchrotron emission \citep{Aharonian2000,Mucke2001} or interact with ambient photon fields via $p\gamma$ interactions. 
The latter produces secondary pions. The charged pion channel proceeds as: 
$$
p + \gamma \;\;\rightarrow\;\; n + \pi^{+},
$$ 
$$
\pi^{+} \;\;\rightarrow\;\; \mu^{+} + \nu_{\mu},  
$$
$$
\mu^{+} \;\;\rightarrow\;\; e^{+} + \nu_{e} + \bar{\nu}_{\mu},  
$$
resulting in high-energy neutrinos accompanied by secondary leptons (electrons and positrons), which can radiate via synchrotron and inverse-Compton processes. In parallel, the decay of neutral pions,

$$
\pi^{0} \;\;\rightarrow\;\; 2\gamma,  
$$
produces very-high-energy $\gamma$-rays. Both the $\gamma$-rays from $\pi^0$ decays and the synchrotron emission of secondary leptons interact with low-energy background photons (e.g., synchrotron photons), leading to $\gamma\gamma$ pair production. The resulting electron–positron pairs can again emit synchrotron radiation, generating fresh high-energy $\gamma$-rays that undergo further absorption. This cycle of $\gamma\gamma$ absorption, synchrotron radiation, and re-absorption initiates an electromagnetic cascade.

As a result, the initially produced high-energy radiation is progressively reprocessed, with the emergent cascade photons accumulating predominantly in the X-ray to MeV band, depending on the specific physical environment (e.g., magnetic field strength, photon density). This tight coupling of hadronic processes with electromagnetic cascades provides a natural framework linking the production of high-energy $\gamma$-rays and neutrinos in blazars.

The first compelling evidence for a blazar–neutrino association so far came with the detection of a $\sim$ 290 TeV IceCube neutrino (event IC-170922A) in spatial and temporal coincidence with a gamma-ray flare from the BL Lac object TXS 0506+056 \citep{IceCube2018}. The association was further supported by archival IceCube data, which revealed a $\sim$3.5$\sigma$ excess of neutrino events from the same direction during a $\sim$6-month period in 2014–2015. This discovery triggered extensive multi-wavelength follow-up campaigns  \citep[e.g.][]{Keivani2018,Ansoldi2018,Cerruti2019} and placed blazars firmly among the most promising candidate neutrino sources. Since then, several studies have reported other possible positional and temporal coincidences between IceCube neutrino events and flaring blazars, although the overall contribution of blazars to the diffuse neutrino flux remains uncertain and may be limited to $\lesssim 30\%$ \citep{Aartsen2017}.

From a theoretical standpoint, efficient neutrino production in blazars requires a combination of factors: (i) the presence of ultra-relativistic protons, (ii) sufficiently dense target photon fields for $p\gamma$ interactions, (iii) high enough proton energies to surpass the pion production threshold, and (iv) a physical environment in which the proton acceleration and interaction timescales are compatible \citep{Murase2014,Petropoulou2015}. FSRQs, with their luminous BLRs and dusty tori, provide strong external photon fields that can serve as effective targets for $p\gamma$ interactions  \citep[e.g.][]{Murase2014,Petropoulou2015}, making them potentially more efficient neutrino emitters than BL Lacs under similar jet powers. However, the same dense photon fields can also lead to strong $\gamma\gamma$ absorption, complicating the observational link between gamma rays and neutrinos.

In this context, any positional and temporal coincidence between a neutrino alert and a blazar flare offers an important opportunity to probe the hadronic content of AGN jets and to test theoretical models of multi-messenger emission. On 2024 January 5 at 12:27:42.57 UT, the IceCube Neutrino Observatory detected the track-like neutrino event IC240105A \citep{gcn35485}. The event’s reconstructed best-fit position is RA = 72.69$^\circ$ and  Dec. = +11.42$^\circ$ (J2000),
with a 90\% point-spread-function (PSF) containment region of (+0.53, -0.33) deg in right ascension and (+0.20, -0.08) deg in declination. 
Within this localization uncertainty, one cataloged $\gamma$-ray source is found: 4FGL J0449.1+112, listed in the Fermi-LAT 4FGL-DR4 catalog and associated with the FSRQ, PKS 0446+11. This source, at a redshift of 2.15, is positioned at RA = 72.28$^\circ$ and Dec. = +11.36$^\circ$ (J2000), lying only 0.4$^\circ$ from the best-fit neutrino position and well within the 90\% containment region. 

The positional coincidence suggests PKS 0446+11 as a potential counterpart to the neutrino event detected on January 5, 2024. Moreover, the source belongs to the population of VLBI-selected, radio-bright AGN identified by \citet{Plavin2023} as statistically correlated with IceCube neutrinos, strengthening its candidacy as a potential neutrino emitter.  PKS 0446+11 is classified as a prototype ``MeV blazar" \citep{ghisellini2009, sbarrato2015, marcotulli2017}, characterized by a synchrotron peak in the far-infrared, a Compton peak in the MeV range, and a high $\gamma$-ray luminosity. An optical emission line at $\sim$4880~\AA ~
identified as C IV $\lambda$1550 yields a redshift of z = 2.153 \citep{shaw2012}, and SED modeling indicates a central black hole mass of $\sim 5 \times 10^{8} M_{\odot}$ \citep{marcotulli2017}.

Following the neutrino alert in January 2024, prompt multi-wavelength follow-up campaigns were triggered \citep{Sinapius24,Eppel24,atel16402_opticalBVR,atel16407_LASToptcal,RATAN_2024,atel16417_NuStar,atel16453_NICER, Stein_35584}, revealing that the source was in an elevated $\gamma$-ray flux state around the time of the neutrino detection in early 2024 \citep{Sinapius24}, following a major flare in November 2023 \citep{Giroletti23}.
\textit{Swift}-XRT follow-up in early January 2024 showed the source in an elevated X-ray flux state compared to archival levels \citep{Prince24}. PKS 0446+11 has also recently been reported to undergo a dramatic change in its optical spectrum, with strong variations in the continuum relative to the C IV emission line flux and equivalent width \citep{PAIANO_S}. This further highlights the highly variable nature of the source across multiple wavebands.

In this paper, we present a detailed multi-wavelength analysis of PKS 0446+11 around the time of IC240105A, with the aim of assessing the plausibility of a neutrino association and constraining the physical processes at play. Using data from \textit{Fermi}-LAT, \textit{Swift}-XRT/UVOT, and archival observations, we construct broadband light curves and an SED spanning radio to gamma-ray energies. We model the SED with both leptonic and lepto-hadronic frameworks, focusing on the ability of each scenario to reproduce the observed electromagnetic emission and to account for the expected neutrino output. In particular, we compare the predicted neutrino spectra from the hadronic model to the IceCube 90\% upper limits \citep{atel16414_upperlimit} for this event, providing constraints on the hadronic contribution to the source’s high-energy emission.

The structure of this paper is as follows. Section 2 presents background about the neutrino event and the follow-up observations. In Section 3, we describe the multi-wavelength data reduction and analysis. Section 4 presents the multi-wavelength light curves and variability analysis, while Section 5 describes the SED modeling and discusses the implications for neutrino production. Section 6 summarizes our conclusions and outlines prospects for future multi-messenger monitoring of PKS 0446+11 and similar sources.

\section{Neutrino Event IC240105A and Multi-wavelength Follow-up} 

On 2024 January 5 at 12:27:42.57 UT, IceCube reported the detection of a track-like high-energy neutrino (IC240105A) with a best-fit position of RA = 72.69$^\circ$, Dec. = +11.42$^\circ$ (J2000) with a 90\% containment radius of $\sim 1-2^\circ$.
Within the 90\% localization region, Fermi-LAT identifies two $\gamma$-ray sources listed in the 4FGL-DR4 catalog \citep{ballet2024}. These are 4FGL J0449.1+1121, associated with the FSRQ PKS 0446+11 (z = 2.15; \citealt{shaw2012}), located 0.4$^\circ$ from the best-fit neutrino position, and 4FGL J0458.0+1152 (NVSS J045804+115142), a blazar of uncertain type, located at a separation of 1.9$^\circ$. A preliminary LAT analysis over the 24 hours preceding the neutrino detection (T0 = 2024-01-05 12:27:42.57 UT) yielded no significant $\gamma$-ray excess ($>$ 5$\sigma$) from either source.

However, on longer timescales, LAT significantly detected only PKS 0446+11. A preliminary LAT analysis showed that the source was already in an elevated $\gamma$-ray state during the weeks before T0. A one-month integration prior to T0, yields a $>$ 5$\sigma$ detection, with a flux of (1.9 $\pm$ 0.3)$\times$10$^{-7}$ ph cm$^{-2}$ s$^{-1}$ (E $>$ 100 MeV; statistical uncertainty only) \citep{Sinapius24}, more than three times higher than its 4FGL-DR4 cataloged average. The source had previously undergone a pronounced $\gamma$-ray flare in November 2023, during which its flux reached $\sim$18 times the catalog average \citep{Giroletti23}, marking it as one of the brightest flaring states ever recorded for this object.  

Swift-XRT follow-up in early January 2024 revealed the source in an elevated X-ray state. On January 6 and 8, 2024, the preliminary analysis yielded flux estimates of 
(4.0 $\pm$ 1.0)$\times$10$^{-12}$ and (4.4 $\pm$ 0.7)$\times$10$^{-12}$ erg cm$^{-2}$ s$^{-1}$, respectively, an order of magnitude higher than the archival level of 2.5$\times$10$^{-13}$ erg cm$^{-2}$ s$^{-1}$
(0.3–10 keV) measured in 2015. The corresponding photon indices of 1.4 $\pm$ 0.5 and 1.21 $\pm$ 0.22 indicate spectral hardening with increasing flux, consistent with a harder-when-brighter trend \citep{Prince24}.

At submillimeter wavelengths, SCUBA2 observations on January 6, 2024 measured fluxes of 1130 mJy at 850 $\mu$m and 838 mJy at 450 $\mu$m \citep{Huang24}, indicating that the source remained in a bright state, but had slightly faded from its peak on November 5, 2023. This maximum, identified in the ALMA calibrator catalog, coincided with the strong $\gamma$-ray flare reported in November 2023 \citep{Giroletti23}, further supporting enhanced jet activity during that period.

At radio frequencies, Effelsberg observations on 7 January 2024 revealed the source in a very high-state, exhibiting an inverted spectrum up to 44 GHz with flux densities (1.1–2.35 Jy) close to its historical maximum-significantly above quiescent levels \citep{Eppel24}.

The source is also monitored by the MOJAVE program at 15 GHz with the VLBA. These observations reveal a remarkable $\sim 90^{\circ}$ swing in the core electric vector position angle (EVPA), confirmed by multiple epochs. A detailed analysis of the polarization behavior will be presented in Kovalev et al. (2025, submitted).

\section{Data reduction and analysis}
Below we summarize the $\gamma$-ray and X-ray/UV data reduction procedures to investigate the multi-wavelength activity of PKS 0446+11 in connection with the IceCube neutrino alert IC240105A. The Fermi-LAT dataset covers a symmetric $\pm$3 month window centered on the neutrino event (2023 October 5 – 2024 April 5), while Swift observations were analyzed for the period available within $\sim$ 1 month following the neutrino event.

\subsection{Fermi-LAT}
The Large Area Telescope (LAT) 
aboard the Fermi satellite, launched by NASA in 2008 \citep{Atwood_2009}, is the primary $\gamma$-ray instrument covering the 20 MeV to over 300 GeV energy range, with a wide field of view of $\sim$ 2.4 steradians and a full-sky scan every $\sim$3 hours. PKS 0446+11 is a known $\gamma$-ray blazar regularly monitored by Fermi-LAT.

We analyzed Pass 8 LAT data spanning 2023 October 5 to 2024 April 5 (MJD 60222-60405).
The analysis was conducted over a 10$^\circ$ radius region of interest (ROI) centered on the radio position of PKS 0446+11, following standard Fermi-LAT procedures described in the Fermi Science Support Center (FSSC) analysis threads \footnote{\url{https://fermi.gsfc.nasa.gov/ssc/data/analysis/documentation/}}. Good-time intervals were selected with the standard filter ``(DATA\_QUAL\>0)\&\&(LAT\_CONFIG==1)", retaining both front and back events, and adopting `evclass=128' and `evtype=3'. 
 Events with zenith angles $> 90^\circ$ were excluded to reduce contamination from the Earth’s limb. Diffuse backgrounds were modeled with the Galactic model ``\emph{$gll_-iem_-v07$}" and the isotropic model ``\emph{$iso_-P8R3_-SOURCE_-V3_-v1$}", with instrument response functions P8R3\_SOURCE\_V3 applied throughout. The model XML included all fourth Fermi-LAT catalog sources (4FGL; \citealt{Abdollahi_2020}) within 15$^\circ$ of the region of interest (ROI) center. Spectral models and source parameters were optimized using the maximum likelihood method, with the detection significance evaluated through the test statistic (TS). Sources with TS $<$ 9 (corresponding to $\sim$3$\sigma$ significance; \citealt{Mattox_1996}) were excluded from further analysis.

We performed an unbinned likelihood analysis for spectral fitting and binned for light-curve generation. Spectral parameters of PKS 0446+11 and the normalizations of nearby sources with TS $\ge$ 10 were left free, while all other sources were fixed to their catalog values. Sources with TS $<$ 10 were iteratively removed from the model. The default spectral model for PKS 0446+11 was a log-parabola, with parameters optimized via likelihood analysis. The integrated spectrum over 0.1–300 GeV for the selected period was fitted with this model.

For lightcurve production, all background/source parameters except those of PKS 0446+11 were frozen, and time-binned (6-hour) fluxes and spectral parameters were extracted. In SED binning, 95\% upper limits were reported for bins with TS $<$ 9 \citep[approximately 3$\sigma$ significance;][]{Mattox_1996}, following established LAT analysis practice.

\subsection{Neil Gehrels Swift Observatory}
The Neil Gehrels Swift Observatory \citep{Gehrels_2004} is a multi-wavelength mission equipped with three instruments, such as the Ultraviolet / Optical Telescope (UVOT), the X-Ray Telescope (XRT) and the Burst Alert Telescope (BAT), covering optical-UV to hard X-ray energies. PKS 0446+11 has been observed with Swift through both routine monitoring and target-of-opportunity (ToO) programs using XRT and UVOT. The source was in a highly active state in the months preceding the neutrino event, including a major $\gamma$-ray flare in November 2023. While no contemporaneous Swift observations were available during this $\gamma$-ray outburst, follow-up XRT/UVOT observations in early January 2024 showed that the source remained in an elevated X-ray flux state compared to archival measurements \citep{Prince24}, confirming its prolonged high activity. For this study, we analyzed all available Swift observations for the period from January 6 to February 11, 2024 (MJD:60315.80-60351) within the time window considered for the Fermi-LAT analysis.

\subsubsection{XRT} 

The XRT operates in the 0.3–10 keV energy band. We retrieved all Photon Counting (PC-mode) observations from HEASARC within the analysis window and processed them using the standard XRTPIPELINE with the CALDB version 0200305. Source and background spectra were extracted from circular regions; for typical, non–pile-up exposures we adopted radii of $\sim$20 arcsec for the source and $\sim$40 arcsec for the background. For observations affected by pile-up, identified when the count rate exceeded 0.5 ct s$^{-1}$, we used annular source regions (inner/outer radii $\sim$4 arcsec/20 arcsec). Ancillary response files (ARFs) were generated with XRTMKARF and appropriate redistribution matrices (RMFs) were applied. Spectra were grouped with grppha to a minimum of 20 counts per bin and fit in XSPEC with an absorbed power-law, fixing the neutral hydrogen column to the Galactic value \citep{nH_2016} toward PKS 0446+11. The resulting fluxes and photon indices were used to build the XRT light curve. For the SED modeling, all Swift-XRT observations obtained between January 5 and February 11 were combined to produce a time-averaged spectrum representative of the elevated state during this period.

\subsubsection{UVOT}

The UVOT \citep{Roming_2005} operates with three optical filters (V, B, U) and three ultraviolet filters (W1, M2, W2). For this work, UVOT image files were processed with the UVOTSOURCE task. Source counts were extracted using a circular aperture of 5 arcsec radius centered on PKS 0446+11, while background counts were taken from a nearby 10 arcsec source-free region. The observed magnitudes were corrected for Galactic extinction using the reddening value of E(B-V) = 0.4255 mag from \cite{SF11} and the standard extinction ratios $A_V$/E(B-V) for each filter were taken from \cite{Giommi_2006}. Finally, extinction-corrected magnitudes were converted to flux densities using the updated UVOT photometric zero-points of \citet{Breeveld2011} and the count-to-flux conversion factors were taken from \citealt{Giommi_2006}. For the SED modeling, the UVOT flux densities from observations between January 5 and February 11 were combined to obtain an averaged optical spectrum contemporaneous with the XRT and LAT data.

 \begin{figure}
       \includegraphics[scale=0.3]{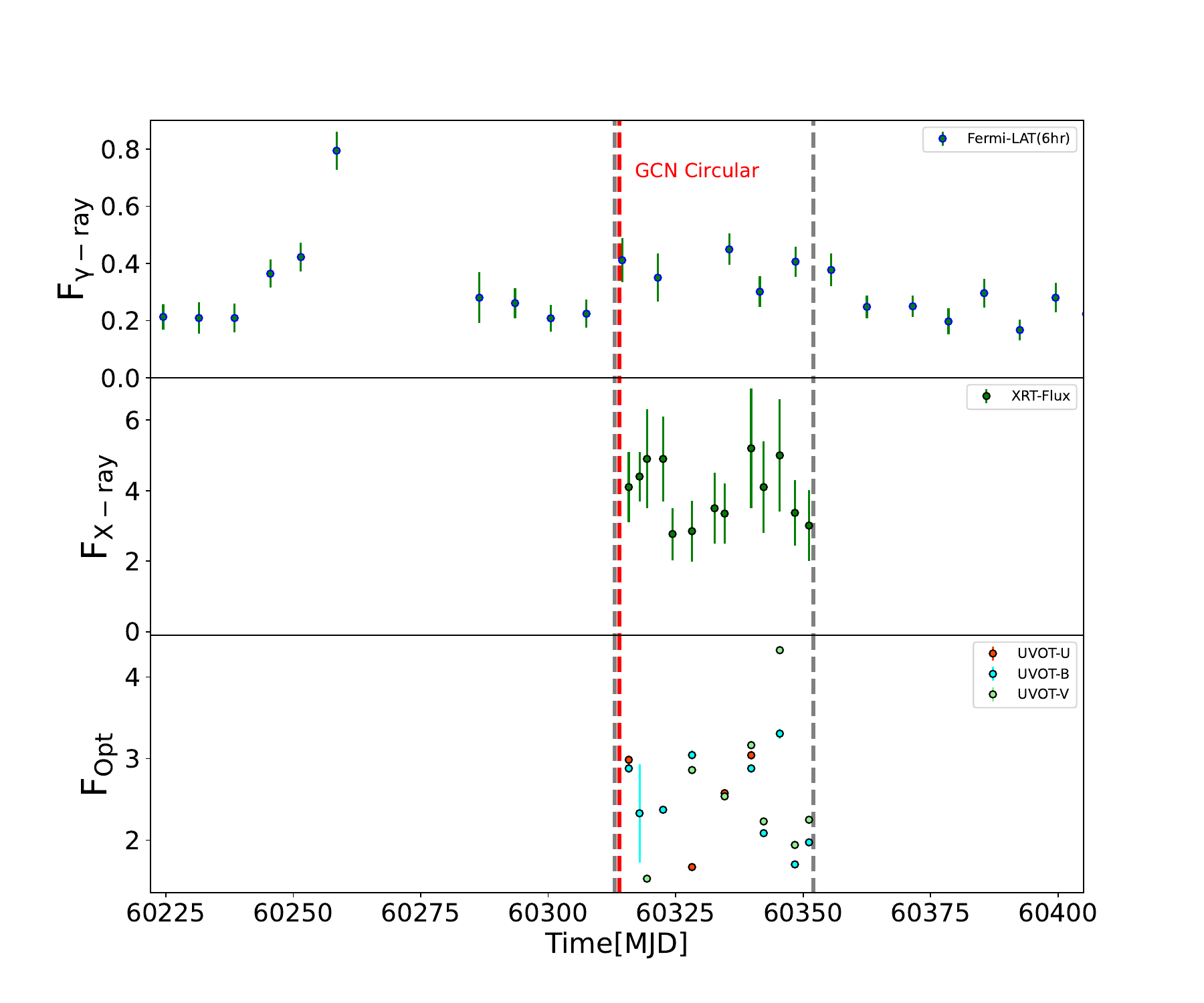}\label{fig:lc}
 \caption{Multi-wavelength light curve of PKS 0446+11 spanning over MJD 60222–60405. The red dashed vertical line marks the detection time of the IceCube neutrino event IC240105A. The time interval bounded by the two grey dashed lines (MJD 60315.80–60351) denotes the period selected for broadband SED modeling
The $\gamma$-ray fluxes are shown in units of 10$^{-6}$ ph cm$^{-2}$ s$^{-1}$
, the Swift-XRT fluxes in 10$^{-12}$ erg cm$^{-2}$ s$^{-1}$, the Swift-UVOT fluxes flux in 10$^{-12}$ erg cm$^{-2}$ s$^{-1}$.}
	   \label{fig:lc}
\end{figure}

\section{Multi-wavelength lightcurves}

In panel 1 of Figure \ref{fig:lc}, the weekly binned $\gamma$-ray light curve of PKS 0446+11 spans from 2023 October 5 to 2024 April 5 (MJD 60222-60405). During late 2023,  the source exhibited pronounced variability, including a strong flare that peaked on November 10, 2023 (MJD 60258), reaching one of the brightest recorded $\gamma$-ray states of the source. By early January 2024, around the time of the IceCube neutrino detection (red dashed line), the $\gamma$-ray flux had decreased, but remained above its long-term baseline, indicating the source was still in an elevated state.

Panel 2 presents the X-ray light curve from Swift-XRT. Although no observations were taken during the November $\gamma$-ray outburst, follow-up monitoring in January 2024 demonstrated that the source was in a high X-ray state compared to archival measurements \citep{Prince24}.

Panel 3 shows the optical light curves from Swift–UVOT in the U, B, and V filters. Although the UVOT sampling is limited, the source exhibits moderate but statistically significant optical variability during early 2024, with flux changes exceeding the typical photometric uncertainties ($\leq$2–3\%) by more than an order of magnitude. This optical activity is contemporaneous with the elevated high-energy state, strengthening the case that PKS 0446+11 remained active across multiple wavebands during and after the neutrino detection.


\subsection{Variability study:}

We estimated the minimum flux doubling/halving timescale, which describes the shortest interval over which the flux changes by a factor of two between successive measurements \citep{Zhang_1999}:

\begin{equation}
t_{d} = \frac{(f_{1}+f_{2})(t_{2}-t_{1})}{2(f_{2}-f_{1})},
\end{equation}

where $t_{1}$ and $t_{2}$ are consecutive observation times with corresponding fluxes $f_{1}$ and $f_{2}$. The minimum value of $t_{d}$ across the light curve is taken as the characteristic variability timescale, $t_{\rm var}$.

Using the 1-day binned $\gamma$-ray light curve, we estimated the minimum variability timescale to be about 1.0 day, based on flux changes from 5.11 $\pm$ 1.3 ($\times10^{-7}\,ph\,cm^{-2}\,s^{-1}$) on MJD 60256.5 to 1.02 $\pm$ 0.23 ($\times10^{-6}\,ph\,cm^{-2}\,s^{-1}$) on MJD 60257.5. The size of the emitting region can be constrained by
R $\lesssim$ c t$_{\rm var} \delta/(1+z)$
where c is the speed of light, t$_{\rm var}$ is the fastest variability timescale, and $\delta$ is the Doppler factor. For t$_{\rm var}$ $\sim$ 1 d, $\delta$ = 35, and redshift z = 2.15, we obtain R $\sim$ 3 $\times 10^{16}$ cm, consistent with the value adopted in our SED modeling. The distance of the emission region from the central engine can be estimated as
d $\sim$ 2 c t$_{var}$ $\delta^2/(1+z)$ \citep{Abdo2011}, which yields d $\sim 2.0 \times 10^{18}$ cm $\sim$ 1 pc, again consistent with the distance used in our modeling.


\begin{figure}
   \includegraphics[scale=0.5]{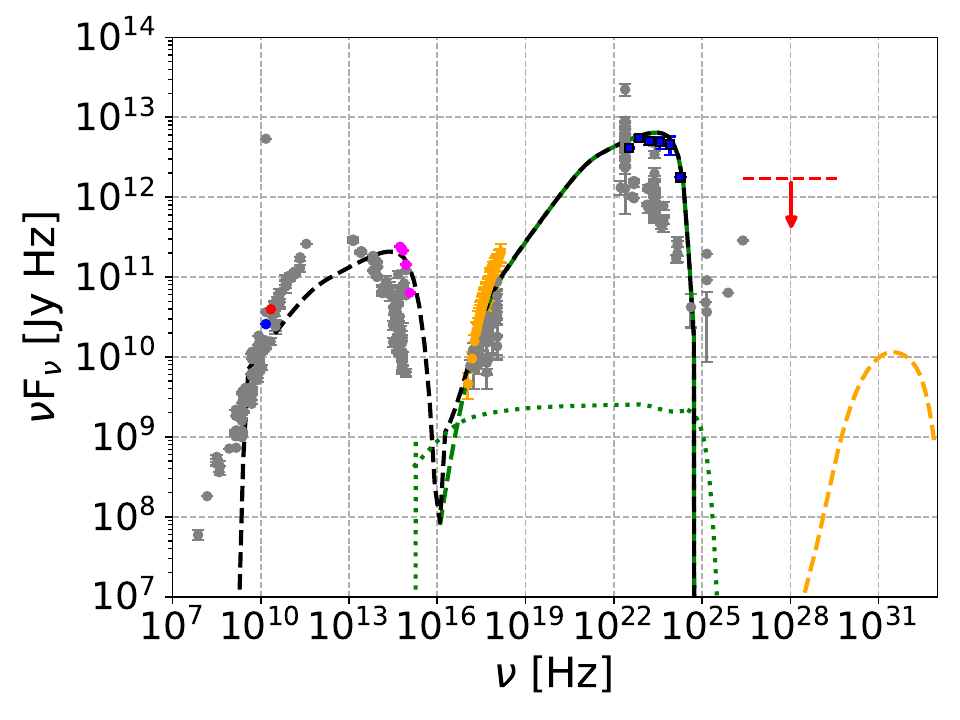}
    \caption{Broadband SED of PKS 0446+11 constructed using multiwavelength observations from RATAN-600 and MOJAVE (radio, red and blue), Swift-UVOT (optical/UV, magenta), Swift-XRT (X-ray, orange), and Fermi-LAT ($\gamma$-ray, blue), with archival data shown in grey. The SED is modeled with a one-zone leptonic emission component (black dashed line) and a sub-dominant hadronic cascade contribution (green dotted line), as described in \S 5. The red downward arrow indicates the 90\% C.L. IceCube muon-neutrino flux upper limit (ATel 16414), and the orange dashed curve represents the model-predicted neutrino spectrum.}
    \label{fig:sed}
\end{figure}

\section{SED modeling}

To study the spectral energy distribution (SED) of PKS 0446+11, we selected the period from January 5 to February 11, 2024 (MJD 60315.80-60351), indicated by the two grey vertical lines in Figure \ref{fig:lc}. This interval was chosen because it directly follows the IceCube neutrino event IC240105A (marked by the red vertical line) and offers contemporaneous multi-wavelength coverage, including $\gamma$-ray, X-ray, and optical observations. The broadband SED corresponds to the time-averaged Swift-XRT and UVOT spectra and the monthly integrated LAT flux over January 5-February 11, providing a representative description of the source's elevated emission state during the month following the neutrino alert. In addition to these observations, we included the MOJAVE 15 GHz VLBA flux density measurement from 2023 December 15 \citep{Lister_2018} and the 22 GHz flux density measurement from RATAN-600 reported in January 2024 \citep{RATAN_2024}, as both epochs are close to the selected time window for our SED modeling.
The SED is presented in Figure \ref{fig:sed}.
The proximity of this time window to the neutrino detection maximizes the relevance of the SED modeling for probing possible multi-messenger connections. We modeled the SED using a stationary, single-zone framework, incorporating both leptonic and hadronic scenarios. For this purpose, we adopted the numerical code developed by \cite{Bottcher2013}, which self-consistently treats the injection, cooling, and radiation of relativistic particles in the jet, as well as the associated radiative transfer. The key elements of the model are summarized below.

\subsection{Stationary Leptonic Emission Model:} 

In the leptonic scenario, the broadband SED of PKS 0446+11 is well reproduced by synchrotron radiation from a nonthermal electron population producing the low-energy component, and inverse Compton (IC) scattering off the same electron population producing the high-energy component. We adopted a stationary, homogeneous emission model of \cite{Bottcher2013} in which a population of ultrarelativistic electrons is injected with a power-law distribution into a spherical region of comoving radius $R$. This emission region propagates along the jet with the bulk Lorentz factor $\Gamma$, corresponding to a relativistic speed $\beta_\Gamma c$, and contains a tangled magnetic field of co-moving strength $B$. The electron distribution cools via synchrotron and Compton emission. 

We estimated the size of the BLR using the disk luminosity of $L_{\rm disk} = 10^{46.54}\ {\rm erg\ s^{-1}}$ \citep{marcotulli2017}, which results in $R_{\rm BLR} \sim 5.9 \times 10^{17}$ cm, while the dusty torus is expected at $R_{\rm DT} \sim 5.9 \times 10^{18}$ cm. Thus, the inferred emission region is located near the transition between the outer edge of the BLR and the onset of the dusty torus. Accordingly, our model includes external photon fields from both the BLR and dusty torus. The BLR radiation field is characterized by a blackbody peaking at the Ly$\alpha$ line with $T = 4.5\times10^4$ K with an energy density of $u_{\rm BLR} \approx 1 \times 10^{-5}\ {\rm erg\ cm^{-3}}$, while the torus field is represented as a blackbody with $T_{\rm DT} = 10^{3}$ K and $u_{\rm DT} \approx 6.8 \times 10^{-5}\ {\rm erg\ cm^{-3}}$. These values indicate that both components contribute to the external Compton scattering, though the torus photons dominate, consistent with the expected physical conditions at that location. The EC emissivity is evaluated using the head-on approximation to the full Klein–Nishina cross section \citep{Dermer2009}, and electron cooling is treated self-consistently. The particle escape timescale is parameterized by $t_{\rm esc} = \eta_{\rm esc} R/c$, where $\eta_{\rm esc}$ is the dimensionless escape timescale parameter.

The code solves for an equilibrium electron distribution by balancing injection, radiative cooling, and escape, iterating until convergence is achieved. This framework yields the synchrotron, SSC, and EC emission components that together reproduce the broadband SED. With the chosen torus and BLR parameters, the isotropic external radiation field approximation provides a good fit to the EC component, where a double EC contribution reproduces the X-ray and $\gamma$-ray emission. In the SED in Figure \ref{fig:sed}, the black dashed line represents the total emission from the leptonic scenario.
Table \ref{sed_parameters} lists the best-fit model parameters used in our lepto-hadronic modeling. The electron distribution extends from $\gamma_{min} \approx 1.1\times{10^{2}}$ to  $\gamma_{max} \approx 4.5\times{10^{3}}$, implying that electrons are accelerated to multi-GeV energies in the comoving frame. The magnetic field B $\approx$ 1.1 G and bulk Lorentz factor $\Gamma$ = 35 are typical of luminous FSRQs and locate the emission region near the outer BLR/inner torus (as discussed above). These parameter values produce a synchrotron peak in the far-IR and an external-Compton dominated high-energy component, consistent with the observed SED shape.
This modeling framework has been successfully applied to several other blazars, including VHE-detected sources such as W Comae \citep{Acciari2009a}, 1ES 0806+524 \citep{Acciari2009b}, PKS 1424+240 \citep{Acciari2010a}, RGB J0710+591 \citep{Acciari2010b}, and 3C 66A \citep{Abdo2011}, as well as high-redshift FSRQs such as PKS 0528+134 \citep{Palma2011}, 4C+01.02 \citep{Schutte2022}, TXS 1508+572 \citep{Gokus2024}, and GB6 B1428+4217 \citep{Gokus2025}, making it a suitable choice for interpreting the SED of PKS 0446+11.

\subsection{Stationary Hadronic Emission Model:} 

In addition to the purely leptonic scenario, we explored a stationary hadronic framework to assess the possible role of relativistic protons in the broadband emission of PKS 0446+11 and the associated neutrino production. In this picture, ultra-relativistic protons contribute to the high-energy emission both through synchrotron radiation \citep{Aharonian2000,MP2000} and photo-hadronic ($p\gamma$) interactions with internal and external photon fields \citep{MB1992}. The proton synchrotron component is calculated using the standard synchrotron emissivity formalism,

\begin{equation}
P_{\rm syn}(\nu) \propto \int N_p(\gamma_p)\, F\!\left(\frac{\nu}{\nu_c(\gamma_p)}\right)\, d\gamma_p ,
\label{Psyn}
\end{equation}
where $N_p(\gamma_p)$ is the proton distribution and $\nu_c(\gamma_p) = (3 \, e \, B/ 4 \, \pi \, m_p \, c) \, \gamma_p^2$ is the critical proton synchrotron frequency. For photohadronic interactions, we adopted the semi-analytical framework of \cite{Bottcher2013}, which employs the analytical templates of \cite{KA2008} to compute the spectra of photons, neutrinos, and secondary $e^\pm$ from $p\gamma$ interactions. The lepto-hadronic model of  \cite{Bottcher2013} self-consistently accounts for secondary emission, including photons from $\pi^0$ decays, synchrotron radiation of secondary $e^\pm$ from charged pion and muon decays, and employs a semi-analytical treatment of synchrotron-supported pair cascades triggered by internal $\gamma\gamma$ absorption in local soft-photon fields.

This stationary semi-analytical treatment reproduces the broad cascade emission shown as the green dotted line in Figure \ref{fig:sed}, spanning the X-ray to $\gamma$-ray bands.
It is obvious that its spectral shape is inconsistent with the observed X-ray -- $\gamma$-ray spectrum. Hence, a model with hadronically dominated high-energy emission is disfavored, but cascade emission at a sub-dominant level, as in the model shown in Figure \ref{fig:sed}, may still contribute to the high-energy emission and produce a non-negligible neutrino flux.

\subsubsection{Neutrino Constraints:}

The hadronic modeling also provides predictions for the associated neutrino emission from PKS 0446+11. The resulting spectrum (orange curve in Figure \ref{fig:sed}) peaks at PeV energies, consistent with the energy range where photo-meson interactions are most efficient. 
However, this peak lies largely outside IceCube’s primary sensitivity range ($\sim$ 1 TeV - 2 PeV).
For comparison, the red marker in Figure \ref{fig:sed} shows the 90\% confidence upper limit on the muon-neutrino flux derived from IceCube observations of IC240105A. This upper limit constrains how bright PKS 0446+11 could have been as a neutrino emitter without being detected. 
The fact that the modeled neutrino flux peaks above IceCube’s most sensitive energy window underscores a key limitation: while the source remains an intriguing spatial and temporal counterpart candidate, current observations favor a predominantly leptonic origin for its broadband $\gamma$-ray emission.

we used the neutrino spectrum predicted by our model to estimate the expected muon-neutrino event rate in IceCube and KM3NeT. The fluxes were transformed into differential number fluxes per unit energy ($\mathrm{GeV^{-1}\ cm^{-2}\ s^{-1}}$) according to

$$
\frac{d\Phi_{\nu,{\rm all}}}{dE} \;=\; \frac{\nu F_\nu \, {\rm [erg \, cm^{-2} \, s^{-1}]}}{(1.602\times10^{-3} \; {\rm erg/GeV}) \, E^2},
$$
where $\nu F_\nu$ is the energy flux and 
neutrino energies are in units of GeV. To account for neutrino oscillations during propagation, the muon-neutrino component was taken as one third of the all-flavor flux, i.e. $d\Phi_{\nu_\mu}/dE = (1/3)\, d\Phi_{\nu,{\rm all}}/dE$ \citep{Ahlers2015}.

The number of muon neutrinos expected to be detected by IceCube during an observing period $\Delta T_{\rm obs}$ was then calculated as

$$
N_{\nu_\mu}^{\rm obs} \;=\; \Delta T_{\rm obs}\; \int_{\epsilon_{\nu_\mu,\min}}^{\epsilon_{\nu_\mu,\max}} 
\frac{d\Phi_{\nu_\mu}^{\rm obs}}{dE}\, A_{\rm eff}(E,\delta)\, dE,
$$
Here, $A_{\rm eff}(E,\delta)$ represents the detector's effective area as a function of energy and source declination. The integration was carried out numerically using the published effective area curves of IceCube \citep{ICECube_2021} and KM3NeT \citep{KM3NeT_2024}.
For IceCube, the IC86-II configuration was adopted, while for KM3NeT, we used the ARCA230 configuration. For the SED period (January 5 – February 11, 2024; $\sim$35.2 days), the expected number of muon-neutrino events is found to be very small: $N_{\nu_\mu}^{\rm obs} \sim 2.85 \times 10^{-4}$ for IceCube and $\sim 6.76 \times 10^{-4}$ for KM3NeT. Scaling these results to longer exposures yields $N_{\nu_\mu}^{\rm obs} \sim 2.96 \times 10^{-3}$ for IceCube over one year and $\sim 2.96 \times 10^{-2}$ over ten years, while for KM3NeT the corresponding values are $\sim 7.01 \times 10^{-3}$ for 1 year and $\sim 7.01 \times 10^{-2}$ for 10 years. These results demonstrate that the modeled neutrino flux is far below the detection threshold of either detector, implying that no neutrino events would be expected from this source during the SED period or even over decade-long observations.

\section{\label{summary}SUMMARY AND DISCUSSION}

In this work, we carried out a multi-wavelength study of PKS 0446+11, along with leptonic and lepto-hadronic SED modeling, motivated by its positional association with the IceCube neutrino event IC240105A on 5 January 2024 (MJD 60314). The source underwent a strong $\gamma$-ray flare in November 2023 \citep[$\sim 18 \times$ its 4FGL-DR4 catalog average;][]{Giroletti23}, but contemporaneous X-ray and optical/UV data are lacking. However, in the neutrino epoch, the source remained active: the weekly binned $\gamma$-ray flux was $4.1\times10^{-7}\,\mathrm{ph\,cm^{-2}\,s^{-1}}$ ($\sim 6 \times$ the catalog average; 5.3$\sigma$ excess), the X-ray flux on MJD 60315 was $\sim 4.1\times10^{-12}\,\mathrm{erg\,cm^{-2}\,s^{-1}}$ ($\sim 10 \times$ archival), and Swift-UVOT measured a bright optical -- UV state with U-band flux density $\sim3.0\times10^{-12}\,\mathrm{erg\,cm^{-2}\,s^{-1}}$. These contemporaneous enhancements indicate that the jet was still in an elevated multi-wavelength state during the neutrino detection, though not at the level of the November 2023 outburst.

The lack of a strict temporal coincidence between the November 2023 $\gamma$-ray flare and the January 2024 neutrino, together with the elevated state observed at the neutrino epoch, highlights the complexity of associating neutrino emission with electromagnetic activity in blazars. In hadronic scenarios, neutrinos are produced via $p\gamma$ interactions, while the accompanying $\pi^{0}$-decay $\gamma$-rays are usually absorbed in compact, magnetized jet regions and reprocessed into X-rays -- soft $\gamma$-rays through synchrotron-supported cascades. Consequently, neutrino production does not necessarily coincide with the brightest $\gamma$-ray states. The extended high state of PKS 0446+11 around the neutrino epoch suggests that the physical conditions required for efficient hadronic interactions --- sustained particle acceleration and dense photon targets --- may still have been present even after the November 2023 outburst. 

Similar temporal offsets between electromagnetic and neutrino emission have been reported for other candidate sources: TXS 0506+056 exhibited a neutrino excess during a $\gamma$-quiet phase \citep{IceCube18,Padovani18}, while the TDEs AT2019dsg \citep{Stein2021} and AT2019fdr \citep{Reusch2022} produced neutrinos months after their electromagnetic maxima. These examples, together with PKS 0446+11, suggest that time delays between photons and neutrinos may be a generic feature of hadronic sources.

The high redshift of PKS 0446+11 ($z \sim 2.15$) adds further complexity. At such distances, $\gamma$-rays above tens of GeV are strongly absorbed by the extragalactic background light (EBL). We accounted for this using the EBL model of \cite{Finke2010}, ensuring realistic intrinsic SED fits. While EBL absorption suppresses observable $\gamma$-rays, neutrinos travel unimpeded, making high-redshift FSRQs potentially important contributors to the diffuse astrophysical neutrino background even when their $\gamma$-ray signatures appear modest.

Our SED modeling shows that the broadband emission is best explained by a leptonic scenario, with synchrotron emission producing the low-energy hump and external Compton scattering of dusty torus and BLR photons accounting for the X-ray to $\gamma$-ray output. The hard X-ray spectrum of PKS 0446+11, together with its steep $\gamma$-ray SED, is well reproduced in this framework. This is consistent with the picture proposed for high-redshift blazars, where EC dominates over SSC in the X-ray band and the $\gamma$-ray spectra appear steep \citep{ajello2016, marcotulli2017}.
A model dominated by hadronic cascade emission is unable to reproduce the observed SED; thus, $p\gamma$ induced cascades, if present, must be sub-dominant. 
The quantitative neutrino constraints derived from the SED further reinforce this interpretation. The expected number of muon-neutrino events is very low of order $10^{-4}$ over the 35-day SED period and $\leq 10^{-2}$ even for decade-long exposures - for both IceCube and KM3NeT. These values imply that the modeled neutrino flux lies far below the detection thresholds of current or near-future instruments, implying that any hadronic contribution to the emission is negligible compared to the dominant leptonic processes.

In conclusion, PKS 0446+11 is unlikely to be the source of IC240105A, but it remains an instructive case study of a high-redshift blazar with multi-wavelength activity around the time of a neutrino event. Our results underscore two key points: (i) neutrino emission in blazars may occur offset from, or hidden during, the most luminous $\gamma$-ray flares, and (ii) leptonic processes dominate the observed emission of PKS 0446+11, with any hadronic contribution constrained to a sub-dominant role. Looking ahead, deeper MeV - X-ray coverage, population studies of high-redshift blazars, and next-generation facilities such as CTAO and IceCube-Gen2 will be crucial to test the neutrino–blazar connection, constrain hadronic emission in luminous FSRQs, and clarify their contribution to the diffuse astrophysical neutrino flux.

\begin{table*}[t]
  \centering
  \caption{{ \footnotesize 
  \label{sed_parameters} Best-fit parameters for the lepto-hadronic SED modeling.}}
{\footnotesize \begin{tabular}{cll}
      \hline
Name & Symbol/units & value \cr
\hline
Minimum Lorentz factor & $\gamma_{min}$ & 1.1e2\cr
Maximum Lorentz factor & $\gamma_{max}$ & 4.5e3\cr
Injection electron spectral index & $\alpha_{e}$ & 1.6\cr
Injection luminosity & $erg\,s^{-1}$ & 7.0e46\cr
Magnetic field &$B$ [G] & 1.1\cr
Bulk Lorentz factor & $\Gamma$ & 35\cr
Emission region size & $R$ [cm] & 3.6e16\cr

Black hole mass & $M_{0}$ & 5.0e8\cr
Characteristic temp. of IR-torus & Kelvin & 1e3\cr
Energy density of the IR photon field & $erg\,cm^{-3}$ & 6.8e-5\cr
Characteristic temp. of BLR-torus & Kelvin & 4.5e4\cr
Energy density of the BLR photon field & $erg\,cm^{-3}$ & 1.e-5\cr
Proton high-energy cutt-off & GeV & 2.5e9\cr
Proton spectral index &$\alpha_{p}$ & 1.9 \cr
\hline
\end{tabular}}
\end{table*}

\begin{acknowledgments}
{\bf Acknowledgements:}
We thank the anonymous referee for constructive feedback that helped improve the manuscript. R.K. and M.B. acknowledge the support provided by the South African Department of Science, Technology, and Innovation and the National Research Foundation through the South African Gamma-Ray Astronomy Programme (SA-GAMMA). 
\end{acknowledgments}

\vspace{5mm}
\facilities{Fermi(LAT), Swift(XRT and UVOT), AstroSat (SXT/LAXPC), NuStar}

\software{Fermitools (\url{https://fermi.gsfc.nasa.gov/ssc/data/analysis/scitools/})\\
\texttt{HEAsoft}-v 6.27\footnote{https://heasarc.gsfc.nasa.gov/lheasoft/download.html}\citep{2014ascl.soft08004N},
\texttt{XSPEC}\footnote{https://heasarc.gsfc.nasa.gov/xanadu/xspec/}\citep{Arnaud_1996},%
\texttt{FTOOLS}\footnote{https://heasarc.gsfc.nasa.gov/ftools/xselect/}\citep{1995ASPC...77..367B}}



\bibliography{main}{}
\bibliographystyle{aasjournal}

\end{document}